\documentclass[fleqn,usenatbib]{mnras}

\usepackage[T1]{fontenc}
\usepackage{rotating}
\usepackage{pdflscape}

\DeclareRobustCommand{\VAN}[3]{#2}
\let\VANthebibliography\thebibliography
\def\thebibliography{\DeclareRobustCommand{\VAN}[3]{##3}\VANthebibliography}

\usepackage{multicol}

\def\ltsima{$\; \buildrel < \over \sim \;$}
\def\simlt{\lower.5ex\hbox{\ltsima}}
\def\gtsima{$\; \buildrel > \over \sim \;$}
\def\simgt{\lower.5ex\hbox{\gtsima}}
\def\cgs{{erg cm$^{-2}$ s$^{-1}$}}

\def\ergs{{erg s$^{-1}$}}
\def\cm2{{cm$^{-2}$}}

\def\xmm{{XMM-{\em Newton}}}
\def\chandra~{{\em Chandra}}

\def\nhgal{{N$_{\rm H}^{\rm Gal}$}}

\def\chandra{{\em Chandra}}

\def\nhgal{{N$_{\rm H}^{\rm Gal}$}}
\def\nh{{N$_{\rm H}$}}

\def\f14{{10$^{-14}$}}
\def\f13{{10$^{-13}$}}
\def\f12{{10$^{-12}$}}
\def\f11{{10$^{-11}$}}
\def\e22{{10$^{22}$}}

\usepackage{graphicx}	

\usepackage{amssymb}	
\usepackage{amsmath}	
\usepackage{longtable}
\usepackage{newtxtext,newtxmath}

\title[X-ray dual AGN ]{The X-ray view of optically selected dual AGN}

\author[A. De Rosa et al.]{
Alessandra De Rosa,$^{1}$\thanks{E-mail: alessandra.derosa@inaf.it.}
Cristian Vignali$^{2,3}$,
Paola Severgnini$^{4}$,
Stefano Bianchi$^{5}$,
Tamara Bogdanovi\'c$^{6}$,
\newauthor
Maria Charisi$^{7}$,
Matteo Guainazzi$^{8}$,
Zoltan Haiman$^{9}$,
S. Komossa$^{10}$, 
Zsolt Paragi$^{11,12}$, 
Miguel Perez-Torres$^{13}$,
\newauthor
Enrico Piconcelli$^{14}$,
Lorenzo Ducci$^{15}$,
Manali Parvatikar$^{15}$,
Roberto Serafinelli$^{1}$
\\
$^{1}$ INAF - Istituto di Astrofisica e Planetologia Spaziali (IAPS), via Fosso del Cavaliere, Roma, I-133, Italy\\
$^{2}$ Dipartimento di Fisica e Astronomia ``Augusto Righi", Universit\`a degli Studi di Bologna, Via Gobetti 93/2, 40129 Bologna, Italy \\
$^{3}$ INAF - Osservatorio di Astrofisica e Scienza dello Spazio di Bologna, Via Gobetti 93/3, 40129 Bologna, Italy\\
$^{4}$ INAF – Osservatorio Astronomico di Brera, via Brera 28, I-20121, Milano, Italy \& via Bianchi 46, I-23807, Merate, Italy\\
$^{5}$ Dipartimento di Matematica e Fisica, Universit\`a degli Studi Roma Tre, via della Vasca Navale 84, 00146 Roma, Italy\\
$^{6}$ School of Physics and Center for Relativistic Astrophysics, Georgia Institute of Technology, Atlanta, GA 30332, USA\\
$^{7}$ Department of Physics and Astronomy, Vanderbilt University, 2301 Vanderbilt Place, Nashville, TN 37235, USA \\
$^{8}$ ESA - European Space Research and Technology Centre (ESTEC), Keplerlaan 1, 2201AZ Noordwijk, the Netherlands \\
$^{9}$  Department of Astronomy, Columbia University, New York, NY, USA\\
$^{10}$ Max-Planck-Institut f{\"u}r Radioastronomie, Auf dem H{\"u}gel 69, 53121 Bonn, Germany\\
$^{11}$ Department of Experimental Physics, University of Szeged, D\'om t\'er 9, H-6720 Szeged, Hungary\\
$^{12}$ Joint Institute for VLBI ERIC, Postbus 2, NL-7900 AA Dwingeloo, The Netherlands\\
$^{13}$ Centro de Estudios de la F\'{\i}sica del Cosmos de Arag\'on (CEFCA), 44001 Teruel, Spain\\
$^{14}$ INAF - Osservatorio Astronomico di Roma, via Frascati 33, 00040 Monte Porzio Catone (Roma), Italy \\
$^{15}$ Institut f{\"u}r Astronomie und Astrophysik, Kepler Center for Astro and Particle Physics, Universit{\"a}t T{\"u}bingen, Sand 1, 72076, T{\"u}bingen, Germany
}

\date{Accepted XXX. Received YYY; in original form ZZZ}
\pubyear{2022}

\begin{document}
\label{firstpage}
\pagerange{\pageref{firstpage}--\pageref{lastpage}}
\maketitle
\begin{abstract}
We present a study of optically selected dual AGN with projected separations of 3--97~kpc. Using multi-wavelength (MWL) information (optical, X-rays, mid-IR), we characterized the intrinsic nuclear properties of this sample and compared them with those of isolated systems.
 Among the 124  X-ray detected AGN candidates, 52 appear in pairs and 72 as single X-ray sources. Through MWL analysis, we confirmed the presence of the AGN in a fraction >80\% of the detected targets in pairs (42 over 52). 
 X-ray spectral analysis confirms the trend of increasing AGN luminosity with decreasing separation, suggesting that mergers may have contributed in triggering more luminous AGN. 
 Through X/mid-IR ratio $vs$ X-ray colors, we estimated a fraction of Compton-thin AGN (with 10$^{22}$ \cm2 $<$ N$_{\rm H} <$10$^{24}$ \cm2) of about 80\%, while about 16\% are Compton thick (CT, with N$_{\rm H}>$10$^{24}$ \cm2) sources. 
 These fractions of obscured sources are larger than those found in samples of isolated AGN, confirming that pairs of AGN show higher obscuration. This trend is further confirmed by comparing the de-reddened [O\ III] emission with the observed X-ray luminosity. However, the derived fraction of Compton-thick sources in this sample at early stage of merging is lower than reported for late-merging dual-AGN samples. Comparing N$_{\rm H}$ from X-rays with that derived from E(B-V) from Narrow Line Regions, we find that the absorbing material is likely associated with the torus or the Broad Line Regions. We also explored the X-ray detection efficiency of dual-AGN candidates, finding that, when observed properly (at on-axis positions and with long exposures), X-ray data represent a powerful way to confirm and investigate dual-AGN systems.
\end{abstract}



\section{Introduction}
\label{sect:intro}
Dual Active Galactic Nuclei (AGN), with separations of  100~pc--100~kpc, have been subject of interest due to their connection with Massive Black Holes (MBHs) triggering through mergers \citep{dimatteoetal05,treisteretal12} and as precursor of MBH coalescence (\citealt[and reference therein]{astroLISA2022}).
Dual AGN are hard to detect (see \citealt{derosaetal19} for a review on observational and theoretical aspects). Different techniques are used to select AGN pairs candidates in different wavebands \citep{Burke-Spolaor2011,comerfordetal12,fuetal2015,Foord2020,Mannucci2022}. Most published samples are sparse and not homogeneous and need to be confirmed through further multi-wavelengths (MWL) observational programs. Dual AGN at early stage of galaxies mergers (1--10~kpc scale) have been identified mostly serendipitously \citep{komossaetal03,Ballo2004, bianchietal08,guainazzietal05,piconcellietal10}. Several studies have been performed at low redshift mainly through large surveys in optical, X-rays, IR \citep{kossetal10,comerfordetal12,liuetal11,riccietal17}.
It has been also shown that higher luminosity AGN are found in interacting systems and average luminosity increases with decreasing separation \citep{hou2020,satyapaletal14,kossetal12,Kocevski2012,silvermanetal11},  suggesting that mergers may trigger MBHs activity \citep{dimatteoetal05,treisteretal12}.
In this scenario of merger-triggered accretion, it is expected that MBHs are completely obscured by gas and dust during the final stage of merging \citep{Hopkins2006}.\\
Both cosmological and numerical simulations have investigated the dynamics and physics of merging MBHs, identifying several parameters that contribute in the activation of both MBHs at the same time (galaxy mass ratio, MBH orbital parameters) and the properties of the environment during the merger \citep{Volonteri2022,Blecha2018,Capelo2017}. 
Although observations have shown that AGN in mergers are characterized by higher obscuration with respect to isolated AGN \citep{riccietal2021,guainazzietal21,Pfeifle2019,derosaetal18,satyapaletal17,riccietal17,derosaetal15,kocevskietal15}, as expected from simulations, it still remains unclear how this obscuration evolves along the merger phase.\\
It is therefore fundamental to identify from theory and simulations the observational signatures to be compared with real observations and track the evolution of MBH through its coalescence.
To make a step forward on this research, we need a statistically significant sample of dual AGN covering a wide dynamical range in spatial separations (1--100~kpc scale). 
SDSS  offers a huge pool of optical data to extract good  dual-AGN candidates through imaging (for spatially resolved systems) and  spectroscopy (used for spatially unresolved systems), the so called double-peaked AGN \citep{Wang2009,Ge2012,Smith2010,Kim2020}. 
Double-peaked technique assumes that each AGN in the system carries its own Narrow Line Region (NLR) tracing the systemic velocity of the AGN as they move in their common gravitational potential. 
However, when the nature of the candidate is verified through MWL  observations, only a tiny fraction of about 2\% can be confirmed  as dual AGN.
This is due to the fact that the signature (i.e., the presence of a doubled-peaked profile) is
not unique, indicating other possible effects originating nearby a single
AGN (e.g., matter outflows); besides, the classification of a galaxy as an AGN is not easy due to extinction  \citep{severgnini2012} and/or the presence of emission from star forming (SF) regions.\\
X-rays represent an efficient technique to detect and confirm accretion-dominated sources such as AGN, even in the case of moderate absorption ({\it i.e.} Compton-thin; column density \nh $\le$~10$^{24}$~cm$^{-2}$).
When used together with other diagnostics, e.g. mid-IR emission, even heavily obscured systems can be detected through their X-ray luminosity.\\
With the main objective of characterizing an homogeneous sample of  dual AGN at kpc separation, and to compare their properties with isolated AGN, in this paper we present the X-ray study of a sample of optically selected dual AGN with separation in the range 3--97~kpc.
We analysed the X-ray archival data from \xmm\ and \chandra\  making use of targets selection in 4XMM \citep{webb2020} and CSC2 \citep{Evans2020,Evans2010}.
The paper is structured as follows. In Sect. \ref{sect:sample} we present 
the sample and its selection, while the MWL data analysis is presented in Sect. \ref{sect:spec analysis}. The diagnostic tools used to identify AGN are shown in Sect. \ref{sect:AGN ident}. We discuss our main results in Sect. \ref{sect:discussion} and provide a summary in Sect. \ref{sect:conclusions}.
Throughout the paper we adopt a concordance cosmology with H$_{0}$ = 70 km\,s$^{-1}$\,Mpc$^{-1}$, $\Omega_\Lambda$= 0.7, $\Omega_M$ = 0.3. Errors and upper limits quoted in the paper correspond to the 90 per cent confidence level, unless noted otherwise.

\section{Sample selection}
\label{sect:sample}

We considered the optically selected (SDSS) dual AGN by \citet{liuetal11}, containing 2488 AGN in pairs (1244 systems at $\bar{z}$ $\sim$ 0.1 with line-of-sight velocity offsets $\Delta$v $<$ 600~km s$^{-1}$ and projected separations $<$ 100 h$^{-1}_{70}$~kpc). Among these targets, 302 fall in the sky regions covered by \xmm\ and/or \chandra\  (considering off-axis positions lower than 10$'$ and 15$'$ for \chandra\ and \xmm,  respectively, and exposure larger than 10~ks); the X-ray detection efficiency will be deeply investigated in the following Sect. \ref{sect:discussion:fraction}.
The optical catalogue was cross-correlated with 4XMM-DR10 \citep{webb2020} considering a match position in a circle with of 5\arcsec\ radius (which considers the \xmm\ telescope position accuracy), obtaining 73 sources. Among these 73 sources, 37 appear as single AGN (7 over 37 with angular separation below 15$''$) and 36 appear in pairs (hence, 18 dual-AGN systems).
The main properties of the 18 AGN pairs are reported in the first block of Table \ref{tab:4xmm-sources}. Due to the \xmm\ spatial resolution and the redshift of the optical sample, the minimum projected separation we are able to explore is  7~kpc.
In order to populate the sample with systems at closer separations, we performed the same selection using \chandra\  observations. We used the CSC2 catalogue  \citep{Evans2020,Evans2010}, obtaining 51 additional targets with projected separation in the range 3--71~kpc. 16 CSC2 targets are in 8 pairs (see second block of Table \ref{tab:4xmm-sources}),  while 35 sources appear as single AGN. The spectral analysis of CSC2 targets has been reported in literature only for few sources. However, to include uniformity in the analysis, we re-analysed all of the sources. \\
The final cross-match between optical and X-ray catalogues returns 124 targets (73 in 4XMM and 51 in CSC2): 52 of them appear in 26 dual systems with 
separations of 3.4--97.2~kpc (see Table \ref{tab:4xmm-sources}). 
We noticed that the mean value and distribution of projected separations  between the X-ray detected targets and the parent population from \citet{liuetal11} are in a good agreement (51 \textit{vs} 58~kpc),  although with a large dispersion (standard deviation of 30~kpc vs. 29~kpc).
\\
For all detected targets, we also searched for  mid-IR information.
In this regard, the AllWISE catalog \citep{Wright2010} provides the magnitudes in the Vega photometric system acquired in the
four observational bands of WISE (W1=3.4~$\mu$m, W2=4.6~$\mu$m, W3=12~$\mu$m, W4=22~$\mu$m). 
To obtain flux densities, we used the conversion as in 
\cite{jarrett2011}, i.e., 
$F_{\nu}[Jy]=F_{\nu_O}\times10^{-(m_{Vega}/2.5)}$, where $F_{\nu_O}$ is the zero magnitude flux density corresponding to the constant that gives the same response as that of Vega. \\
For 32 targets (4XMM: 17, CSC2: 15), we reached a sufficient signal-to-noise (S/N) ratio to enable X-ray spectral analysis.
This threshold was choose  of $\sim$ 100 cts for 4XMM-EP8 in the band (0.2--12 keV), while in the case of \chandra\ we used the spectra released in CSC2 (see details in \citealt{Evans2020}). 
In the following section, we explore different diagnostics using a multi-wavelength approach in order to confirm the AGN nature for the 26 dual-AGN systems optically selected. The properties of the optically dual AGN detected as single X-ray sources will be investigated in a forthcoming paper (Parvatikar et al. in preparation).

We note that the dual AGN sample contains both, Seyfert galaxies (or quasars at higher luminosity) and low-ionization nuclear emission-line regions (LINERs), such that pairs are of the types Seyfert--Seyfert, Seyfert--LINER, and LINER--LINER. 
LINERs are known to be a mixed class of objects with diverse line excitation mechanisms discussed in the literature, such as old stars, shocks, halos of elliptical galaxies and even intra-cluster gas, as well as forms of AGN excitation (e.g.  
\citealt{heckman1980,shields1992,Ho1993,komossaetal999}).
Therefore, not all LINERs are expected to be bona-fide AGN. Here, we continue to treat all LINERs as candidate AGN, on a careful object-by-object basis including information from all optical emission lines. For sources that lie in the composite region of diagnostic diagrams (see Sect. \ref{sect:opt analysis}) and in the LINER regime, we use the X-ray information to get further clues on the main power source of these systems and then re-assess their possible AGN nature in Sect. \ref{sect:AGN ident}. \\
Finally, we would also like to note that the majority of the pairs in our sample are at large projected spatial separations, i.e. above 10$''$ (see Table \ref{tab:4xmm-sources}). Therefore, the well-known effect of light spillover into nearby fibers due to seeing, that can mimik AGN pairs when in fact only one of them is active \citep{Husemann2020}, is minimal in our sample.

\begin{table*}
\caption{Dual-AGN candidates detected in  4XMM-\xmm\ and CSC2. Numbering in the first column refers to sources shown in Fig. \ref{fig:bpt}. Sources in $^\ddagger$\textit{italic} are not identified as AGN following our diagnostic (see Sect. \ref{sect:AGN ident}).}
\label{tab:4xmm-sources}
\begin{tabular}{l||c|c|r|r|r|r|r|r|r|c}
\hline
  \multicolumn{1}{|c|}{} &
  \multicolumn{1}{|c|}{ObsID} &
  \multicolumn{1}{|c|}{SimbadName} &
  \multicolumn{1}{c|}{RA} &
  \multicolumn{1}{c|}{DEC} &
  \multicolumn{1}{c|}{z} &
  \multicolumn{1}{c|}{$^{(1)}$sep} &
  \multicolumn{1}{c|}{$^{(2)}$rp} &
  \multicolumn{1}{c|}{IAUNAME} &
  \multicolumn{1}{c|}{$^{(3)}$Net counts} &
    \multicolumn{1}{c|}{ref} \\
      \multicolumn{1}{|c|}{} &
      \multicolumn{1}{|c|}{} &
  \multicolumn{1}{|c|}{(SDSS)} &
  \multicolumn{1}{c|}{(deg)} &
  \multicolumn{1}{c|}{(deg)} &
  \multicolumn{1}{c|}{} &
  \multicolumn{1}{c|}{(arcsec)} &
  \multicolumn{1}{c|}{(kpc)} &
  \multicolumn{1}{c|}{(4XMM/2CXO)} &
  \multicolumn{1}{c|}{(counts)} &
    \multicolumn{1}{c|}{} \\
  \hline
    \multicolumn{11}{|c|}{XMM-Newton} \\
\hline
 1& 0761730401 & J015235.35-083236.6 & 28.14729 & -8.5435 & 0.0517 & 58.5 & 58.9 &  J015235.2-083233 & 73$\pm$13  &  \\
 2&  0761730401 & J015235.99-083138.9 & 28.14996 & -8.52747 & 0.0511 & 58.5 & 58.9 &  J015235.9-083139 & 69$\pm$12 &  \\
 3&  0741580501 & J082321.66+042220.9 & 125.84029 & 4.37247 & 0.0311 & 142.8 & 88.7 &  J082321.6+042221 & 8871$\pm$125 &  H20\\
 4&  0741580501 & J082329.87+042332.9 & 125.8745 & 4.3925 & 0.0308 & 142.8 & 88.7 &  J082329.9+042332 & 39$\pm$10 &  H20 \\
 5&  0740620201 & J083157.64+191241.4 & 127.99021 & 19.21153 & 0.0372 & 55.8 & 41.2 &  J083157.6+191241 & 149$\pm$21 & \\
 6&  0740620201 & J083200.51+191205.8 & 128.00212 & 19.20164 & 0.0375 & 55.8 & 41.2 &  J083200.6+191206 & 86$\pm$13 & \\
 7&  0743110701 & J100133.68+033731.1 & 150.39033 & 3.62533 & 0.0437 & 53.8 & 45.3 &  J100133.4+033731 & 314$\pm$48 &  \\
 8&  0743110701 & J100135.80+033647.8 & 150.39921 & 3.61328 & 0.0427 & 53.8 & 45.3 &  J100135.8+033648 & 260$\pm$29 &   \\
 9&  0503600301 & J100230.89+324252.3 & 150.62871 & 32.71456 & 0.049 & 89.3 & 88.1 &  J100230.8+324248 & 71$\pm$15 &    \\
 10& 0503600301 & J100236.54+324224.2 & 150.65225 & 32.70675 & 0.0505 & 89.3 & 88.1 &  J100236.6+324224 & 2359$\pm$93 &   \\
 11&  0146990101 &J102141.89+130550.3 & 155.42454 & 13.09733 & 0.0765 & 67.1 & 97.2 &  J102141.8+130551 & 45$\pm$11 &  H20\\
12& 0146990101 & $^\ddagger$\textit{J102142.78+130656.1} & 155.42829 & 13.11558 & 0.0763 & 67.1 & 97.2 &  J102142.6+130654 & 179$\pm$17 &  H20 \\
13& 0692330501 & $^\ddagger$\textit{J112545.04+144035.6} & 171.43771 & 14.67658 & 0.034 & 73.6 & 49.7 &  J112545.1+144035 & 622$\pm$30 &  H20, TA18\\   
 14& 0692330501 & J112549.54+144006.5 & 171.45646 & 14.6685 & 0.0339 & 73.6 & 49.7 &  J112549.5+144006 & 249$\pm$21 &  H20, TA18\\
 15&  0555060501 & J120157.72+295926.6 & 180.4905 & 29.99075 & 0.072 & 19.1 & 25.9 &  J120157.8+295927 & 62$\pm$11 &   \\
 16&  0555060501 & J120157.88+295945.5 & 180.49117 & 29.996 & 0.0713 & 19.1 & 25.9 &  J120157.8+295945 & 40$\pm$10 &   \\
 17&  0601780601 & J120443.31+311038.2 & 181.1805 & 31.17728 & 0.025 & 61.5 & 31.0 &  J120443.3+311037 & 16517$\pm$137 &  K10 \\
 18&  0601780601 & J120445.19+311132.9 & 181.18833 & 31.19247 & 0.025 & 61.5 & 31.0 &  J120445.3+311130 & 154$\pm$20 &  K10 \\
 19&  0111281601 & J134130.40-002514.3 & 205.37671 & -0.42072 & 0.0713 & 61.5 & 83.6 &  J134130.5-002512 & 30$\pm$9 &   \\
 20&  0111281601 & J134133.37-002432.0 & 205.38904 & -0.40892 & 0.0719 & 61.5 & 83.6 &  J134133.4-002431 & 42$\pm$9 &  \\
 21& 0783520501 & J145627.39+211955.9 & 224.11417 & 21.33222 & 0.0443  & 68.4 & 59.2 &  J145627.4+211956 & $^\dagger$602$\pm$25  &  ADR18 \\
 22&  0783520501 & J145631.35+212030.0 & 224.13067 & 21.34169 & 0.044 & 68.4 & 59.2 &  J145631.3+212030  & $^\dagger$459$\pm$32  &  ADR18 \\
 23&  0721820201 & J145838.58+382727.9 & 224.66075 & 38.45775 & 0.1358 & 32.6 & 78.8 &  J145838.5+382727 & 124$\pm$16  & \\
 24&  0721820201 & J145840.73+382732.7 & 224.66971 & 38.45908 & 0.1367 & 32.6 & 78.8 &  J145840.6+382730 & 73$\pm$16  &\\
 25&  0147210301 & J160501.37+174632.4  & 241.25571 & 17.77569 & 0.033 & 117.3 & 78.1 &  J160501.3+174632 & 77$\pm$14 & H20,  \\
 26&  0147210301 & J160507.88+174527.6 & 241.28287 & 17.75767 & 0.0334 & 117.3 & 78.1 &  J160508.1+174528 & 109$\pm$27 &  H20,  \\
 27&  0783520301 & J162640.93+142243.6 & 246.67054 & 14.37878 & 0.0479 & 54.1 & 51.3 &  J162640.9+142243 & $^\dagger$364$\pm$20 &  ADR18\\
 28&  0783520301 & J162644.50+142250.6 & 246.68546 & 14.38075 & 0.0484 & 54.1 & 51.3 &  J162644.4+142253 & $^\dagger$162$\pm$15 &  ADR18 \\
 29&  0784521201 & J163102.72+394733.0 & 247.76133 & 39.7925 & 0.0289 & 165.8 & 96.0 &  J163102.7+394733 & 345$\pm$28  &\\
 30&  0784521201 & J163103.40+395018.5 & 247.76421 & 39.83847 & 0.0305 & 165.8 & 96.0 &  J163103.4+395015 & 85$\pm$17  &\\
 31& 0673000147 & J221839.91-002402.0 & 334.66633 & -0.40053 & 0.0948 & 30.7 & 54.2 &  J221839.8-002400 & 38$\pm$10  & \\
 32& 0673000147 & $^\ddagger$\textit{J221840.97-002335.5} & 334.67071 & -0.39319 & 0.095 & 30.7 & 54.2 &  J221840.9-002333 & 20$\pm$8 & \\
 33&  0783520101 & J094554.40+423839.9 & 146.47671 & 42.6444 & 0.0745 & 21.3 & 30.3 &  J094554.4+423839 & $^\dagger$21450$\pm$150 & ADR18 \\
 34&  0783520101 & J094554.49+423818.7 & 146.47701 & 42.6385 & 0.0745 & 21.3 & 30.3 &  J094554.4+423839 & $^\dagger$1275$\pm$37 &  ADR18\\
 35&  0783520201 & J103853.29+392151.1 & 159.72204 & 39.36422 & 0.0548 & 40.4 & 42.9 &  J103853.3+392151 & $^\dagger$6534$\pm$80 & ADR18 \\
 36& 0783520201 & J103855.94+392157.5 & 159.73312 & 39.36660 & 0.0548 & 40.4 & 42.9 &  J103855.9+392157 & $^\dagger$182$\pm$14 & ADR18\\

\hline
 \multicolumn{10}{|c|}{Chandra} \\
\hline
 1&  14965 & J090714.44+520343.4 & 136.81021 & 52.06206 & 0.0596 & 7.6 & 8.9 &  J090714.4+520343 & $41.9^{+ 7.2}_{- 6.5}$  & H19,H20\\
 2&  14965 & J090714.61+520350.6 & 136.81087 & 52.06408 & 0.0602 & 7.6 & 8.9 &  J090714.6+520350 & $ 120.9^{+11.6}_{-11.5}$ & H19,H20 \\
 3&  4110, 4934& J121345.92+024838.9 & 183.44146 & 2.81083 & 0.073 & 3.4 & 4.8 &  J121345.9+024838 & $11.3^{+ 4.}_{- 3.4}$ & H20,I11,R21\\
 4& 4110, 4934 & J121346.07+024841.4 & 183.44212 & 2.8115 & 0.0731 & 3.4 & 4.8 &  J121346.0+024841 & $14.9^{+ 4.5}_{- 3.9}$ & H20,I11,R21\\
 5& 2043 & $^\ddagger$\textit{J124610.10+304354.9} & 191.54212 & 30.73192 & 0.0219 & 37.2 & 16.4 &  J124610.0+304355 & $18.3^{+ 5.4}_{- 4.7}$ & H20\\
 6&  2043 & J124611.24+304321.8 & 191.54683 & 30.72275 & 0.0218 & 37.2 & 16.4 &  J124611.2+304321 & $70.6^{+ 9.0}_{- 9.0}$ & H20\\
 7&  $\clubsuit$ & $^\ddagger$\textit{J125929.96+275723.1} & 194.87483 & 27.95644 & 0.0227 & 71.3 & 34.4 &  J125929.9+275723 & $ 154.2^{+31.3}_{-31.5}$ & H20\\
 8&  $\clubsuit$ & $^\ddagger$\textit{J125934.12+275648.6} & 194.89217 & 27.94683 & 0.024 & 71.3 & 34.4 &  J125934.1+275648 & $334.7^{+30.5}_{-30.2}$ & H20\\
 9& 12242 & $^\ddagger$\textit{J131513.87+442426.4} & 198.80779 & 44.40736 & 0.0354 & 50.9 & 35.9 &  J131513.8+442426 & $14.5^{+ 4.5}_{- 3.9}$ & H20\\
 10&  12242 & J131517.26+442425.5 & 198.82196 & 44.40711 & 0.0355 & 50.9 & 35.9 &  J131517.3+442425 & $3835.5^{+65.4}_{-64.8}$ & H20\\
 11& 2044 & J133817.27+481632.1 & 204.57196 & 48.27564 & 0.0278 & 11.5 & 6.4 &  J133817.3+481632 & $95.2^{+10.8}_{-10.6}$ & Ma12,H20,I20\\
 12& 2044 & J133817.77+481640.9 & 204.57404 & 48.27808 & 0.0277 & 11.5 & 6.4 &  J133817.8+481640 & $205.0^{+15.3}_{-15.2}$ & Ma12,H20,I20\\
 13& 11845 & $^\ddagger$\textit{J150457.12+260058.4} & 226.238 & 26.01625 & 0.054 & 61.6 & 65.2 &  J150457.1+260058 & $68.0^{+ 9.5}_{- 9.4}$ & H20\\
 14& 11845 &  $^\ddagger$\textit{J150501.22+260101.5} & 226.25508 & 26.01708 & 0.0545 & 61.6 & 65.2 &  J150501.2+260101 & $15.1^{+ 4.6}_{- 4.0}$ & H20\\
 15&  14968 & $^\ddagger$\textit{J154403.45+044607.5} & 236.01437 & 4.76875 & 0.042 & 4.1 & 3.4 &  J154403.4+044607 &  $5.0^{+ 2.9}_{- 2.2}$ & H19,H20\\
 16& 14968 & J154403.66+044610.0 & 236.01529 & 4.76947 & 0.0416 & 4.1 & 3.4 &  J154403.6+044609 & $59.3^{+ 8.2}_{- 8.1}$ & H19,H20\\
\hline\end{tabular}

$^{(1)-(2)}$ Angular and projected separation between the sources; $^{(3)}$ Net observed count in (0.2--12 keV) for 4XMM (EP8) and  (2--8 keV) for CSC; 
$^{(4)}$ References: H19: \citealt{Hou2019}; H20: \citealt{hou2020}; K10: \citealt{kossetal10}; ADR18: \citealt{derosaetal18}; I20: \citealt{Iwasawa2020}; MA12: \citealt{Mazzarella2012}; R21:\citealt{riccietal2021}; I11 \citealt{iwasawaetal11}; TA18: \citealt{torres-alba_etal2018}. $^\dagger$ For these sources we report here the values obtained in \citealt{derosaetal18}. 
$\clubsuit$ ObsID=9714, 10672, 13993, 13994, 13995, 13996, 14406, 14410, 14411, 14415.
\end{table*}

\section{Spectral analysis}
\label{sect:spec analysis}
In this section, we present the MWL spectral analysis of our final sample comprising 26 dual AGN candidates (see Table \ref{tab:4xmm-sources}).
The main objective of this study is to confirm the nature of the selected targets and characterize their intrinsic properties such as X-ray luminosity, optical over X luminosity ratio and absorption properties (at nuclear and galaxy scale). As illustrated in Sect. \ref{sect:intro}, X-rays and mid-IR observations, along with optical spectroscopy, provide fundamental diagnostics to detect accreting MBHs even in heavily obscured systems, as expected in dual AGN.

\subsection{Optical analysis}
\label{sect:opt analysis}
 
\begin{figure*}
\includegraphics[width=1\textwidth]{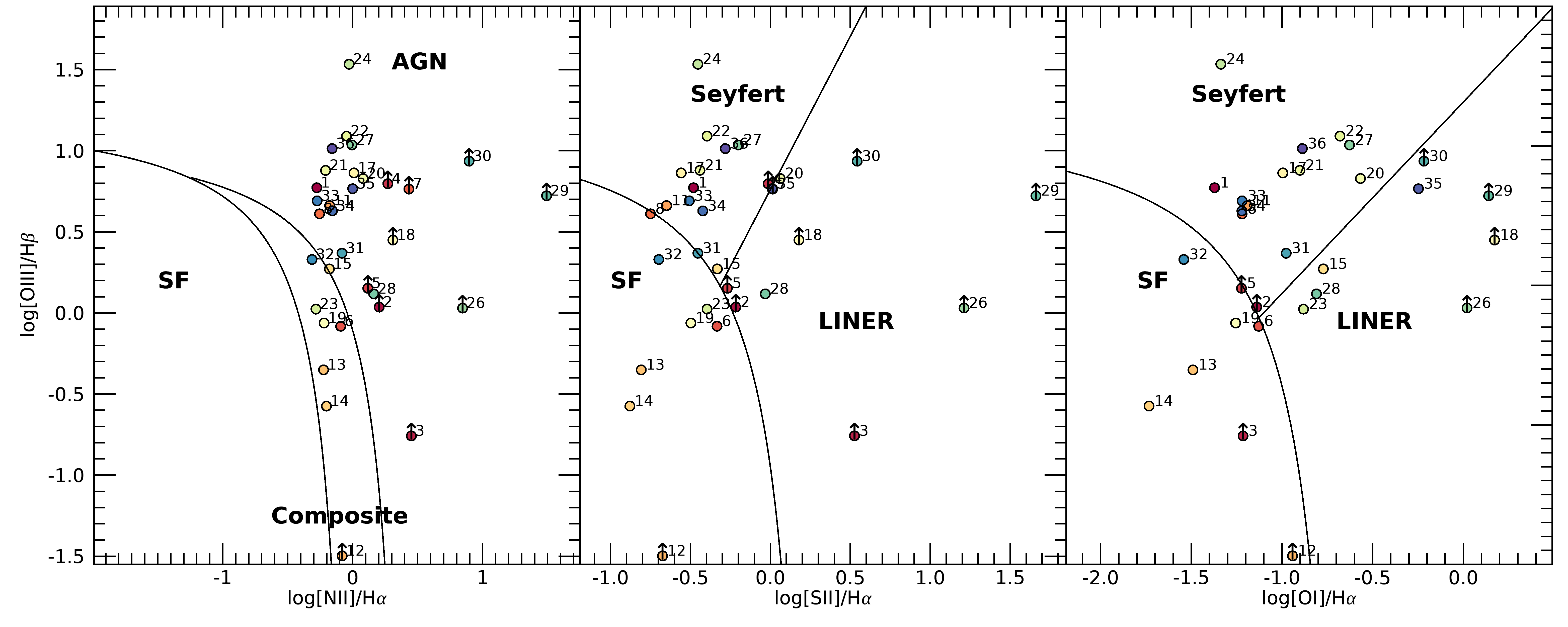} 
\includegraphics[width=1\textwidth]{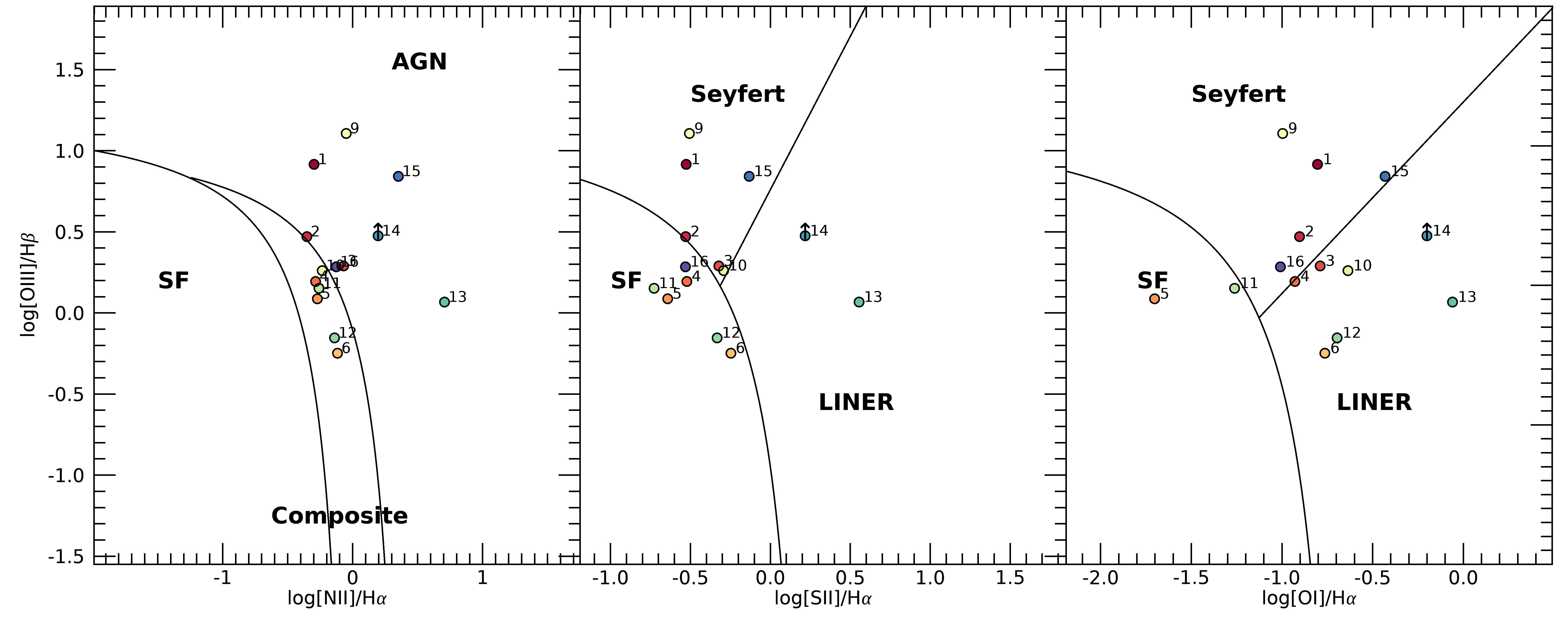} 
\caption{Optical emission-line ratios as classical diagnostic plot for identification of dual-AGN candidates detected in 4XMM (upper panels) and CSC2 (lower panels). Same color refers to members of the same pair. Number of each source refer to Table \ref{tab:4xmm-sources}. AGN, LINER and SF regions in the plots are separated according to \citet{Kewleyetal06}.} \label{fig:bpt}
\end{figure*}

We retrieved the SDSS-III DR12 spectra \citep{SDSS-DR12} at the location of the dual-AGN systems, as reported in Table \ref{tab:4xmm-sources}, from the survey webpage\footnote{http://skyserver.sdss3.org/}. 
The full-band optical spectra has been analysed through the software package QSFit 1.3.0 \citep{qsfit}. This tool automatically takes into account the emission form both AGN and host galaxy, along with a number of broad and narrow emission lines. 
In particular, we extracted the intensities of the primary diagnostic narrow emission lines such as H$\beta$, [OIII] $\lambda5007$, [OI] $\lambda6300$, H$\alpha$, [NII] $\lambda6583$ and [SII] $\lambda\lambda6717,6730$. Line fluxes were corrected for  Galaxy extinction using NED database\footnote{https://ned.ipac.caltech.edu/Documents/References/ExtinctionCalculators}.
The line flux ratios were then plotted in the diagnostic diagrams \citep[BPT diagram]{kauffmannetal03} shown in Fig.~\ref{fig:bpt} (upper panels refers to \xmm\ sources and bottom panels to \chandra\ sources), where the regions  populated with Seyfert galaxies, SF galaxies, and LINERs are identifies as in \citet{Kewleyetal06}. Each number in the plot refers to sources in Table \ref{tab:4xmm-sources}; same colors refer to members of a pair.
Some of the optical spectra are dominated by galaxy contribution, however when possible we also  evaluated the extinction E(B-V) considering the Balmer decrement from the narrow emission lines.
Sources with at least one BPT diagram indicating an AGN or LINER are included in our sample. There are, however, targets whose optical analysis does not confirm their AGN nature. 
We kept these sources in the catalogue and will investigate their classification using further diagnostics such as the X-ray luminosity and mid-IR $vs$ X-ray colors in Sect. \ref{sect:AGN ident}.
We note that only two sources show clear evidence of broad optical lines (J094554.4+423840 and J103853.2+392151, \citealt{derosaetal18}), while in the rest of the sample the galaxy contribution prevent us to detect any strong broad-line component.

\subsection{X-ray analysis}
\label{sect:X-analysis}

We reduced \xmm\ data using SAS software V17.0 with standard settings and the most updated calibration files available at the time of the data reduction. Period of high and flaring background were removed applying appropriate threshold on single events extracted above 10 keV (about 0.4 and 0.35 counts per second in EPIC-pn and EPIC-MOS, respectively).

Depending on the source counts and the separation of the two sources in each system, we extracted EPIC spectra from circular regions with radii in the range 11$''$--25$''$ ; these regions include $\sim$60--90\% of the source counts at 1.5~keV in the EPIC cameras. Background spectra were extracted in the same CCD chip using circular regions free from contaminating sources.
Due to the higher net counts in the 0.3–10 keV energy band with respect to EPIC/MOS, we reported the EPIC/pn spectra analysis and checked a posteriori that pn+MOS data do not improve the fit neither the spectral parameter constraints.
\chandra\ spectra and corresponding response matrices were retrieved from CSC2. When multiple observations of the same source were available, we verified the absence of any sign of variability and the spectra from the individual observations were merged to increase the S/N ratio (the ObsID for each observation is reported in Table \ref{tab:4xmm-sources}).

Among the 52 sources, we were able to extract spectral information for 32 targets (4XMM: 17, CSC2: 15). 
For these sources, EPIC-pn and ACIS background-subtracted spectra were fitted with the software xspec v12.10 \cite{xspec1996} using  Cash statistics \citep{cash1979, wachter1979}.
We applied a phenomenological model composed of (1) an absorbed power-law; this is the nuclear emission  due to Comptonization of electrons in a hot corona by seed photons, probably originated in the accretion disc (e.g. \citealt{haardt&maraschi93,haardtetal94}), (2) a soft, unabsorbed power-law component reproducing the extended soft X-ray emission below $\sim$2~keV, likely associated with the NLR \citep[and references therein]{bianchietal2019}, scattering of the primary emission \citep{uedaetal07}, or to star-formation activity \citep{ranallietal03}.
The baseline model can be written as 
e$^{\rm {-N_{\rm H}^{\rm Gal}\sigma}}$ $(e^{-N_{\rm H}\sigma} K_h E^{-\Gamma_h}+K_s E^{-\Gamma_s})$,
where $\Gamma_h$ and $\Gamma_s$ are the photon indices of the nuclear primary component and soft X-ray component, respectively; $N_{\rm H}^{\rm Gal}$ is the Galactic column density, {N$_{\rm H}$} the cold absorption column density at the redshift of the source; $\sigma$ is the photo-absorption cross-section from \citet{balucinska&mccammon92}.
This simple model is sufficient to get constraints on the main spectral parameters of interest, such as N$_{\rm H}^{\rm z}$, $\Gamma$ and the unabsorbed luminosity.

If count statistic does not allow us to constrain simultaneously $\Gamma_h$ and $\Gamma_s$, we put $\Gamma_h$=$\Gamma_s$, as expected in a scattering model for the soft excess discussed above. We also fixed $\Gamma_h$ to a typical value of 1.9 \citep{riccietal17} when counts statistic does not allow to fit the parameter. The results of the spectral analysis and modeling performed on the sample are reported in Table \ref{tab:xfit} . 

\section{AGN identification}
\label{sect:AGN ident}

We used different diagnostics in order to confirm the AGN nature of our sample against SF galaxies that could produce the observed X-ray luminosity.

We anticipate that, at the end of our MWL analysis, we selected 42 confirmed AGN among 52 candidates. For 24 of these AGN we could perform a detailed X-ray  spectral modeling as described in Sect. \ref{sect:X-analysis}. This "clean sample" of 24 AGN has been then used to calibrate MWL relations in order to get information on the nature (AGN, no-AGN) and absorption properties of the whole sample of 42 dual AGN. Below we report the details of this study.

\begin{figure*}
\centering
\includegraphics[width=0.45\textwidth]{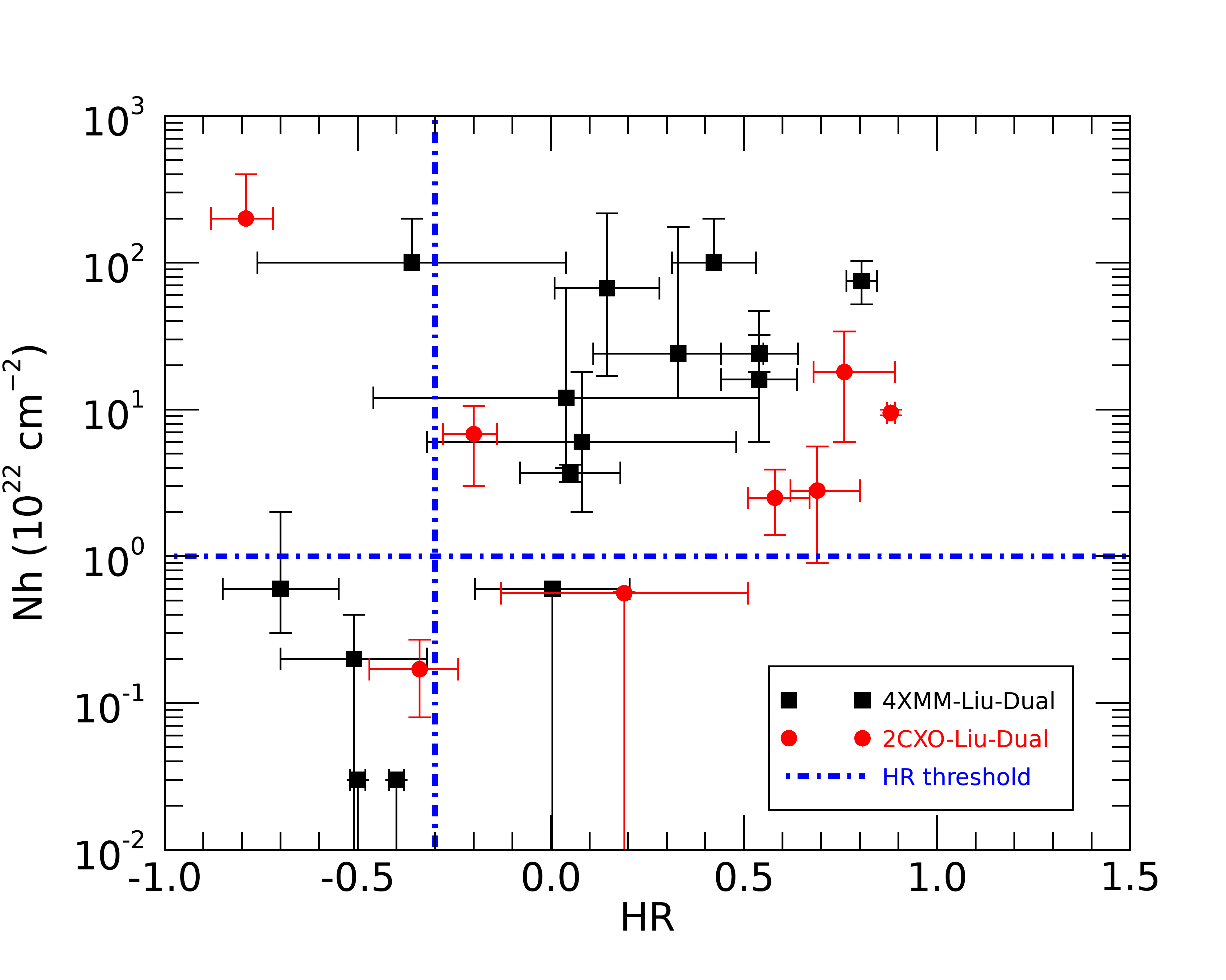}
\includegraphics[width=0.45\textwidth]{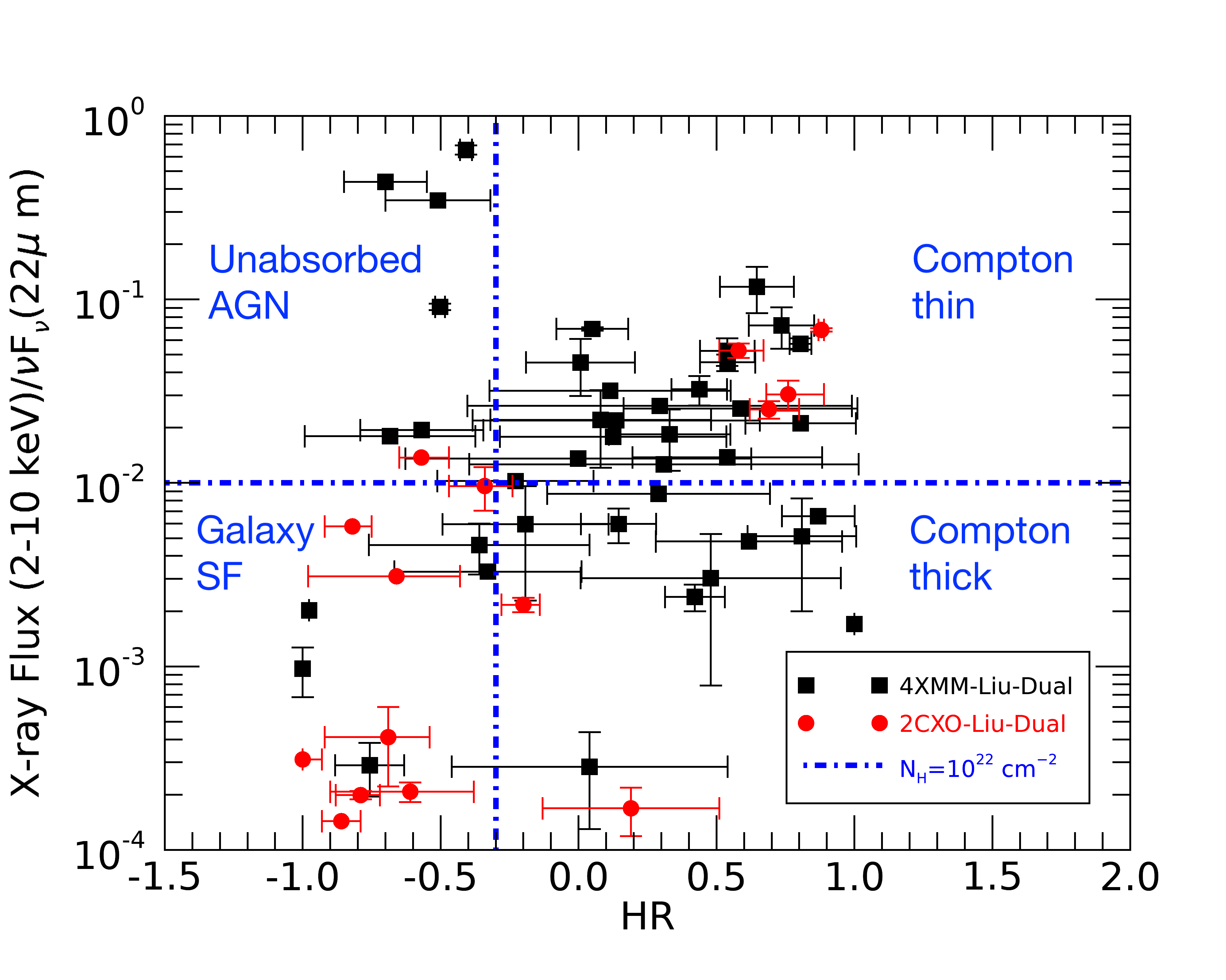} 
 \caption{Left: Absorption column density $vs$ hardness ratio, HR=(H-S)/(H+S) for the clean sample of 24 dual AGN with spectral information (see Table \ref{tab:xfit}). The HR is evaluated considering S=2--4.5 keV and H=4.5--10 keV for 4XMM data, and S=0.5--2 keV and H=2--8 keV for CSC2 data. Dash-dotted vertical line define the threshold for sources with N$_{\rm H}$>10$^{22}$ \cm2 in our sample. Right: AGN identification and absorption diagnostic using mid-IR $vs$ HR for the total sample of 42 confirmed dual AGN (sse Table \ref{tab:4xmm-sources}). Dash-dotted lines identify the regions in the plot where different classes of X-ray and mid-IR emitters are located \citep{severgnini2012}.}\label{fig:nhxir_vs_hr4}
\end{figure*}

As anticipated in the previous section, we retain in the sample targets indicating the presence of an AGN or LINER in at least one optical BPT diagram (see Fig. \ref{fig:bpt}). 
The X-ray luminosity of the sources associated to LINER through optical analysis suggests that their emission is powered by the AGN (see Table \ref{tab:xfit}). 
In fact, a systematic study of a large sample of LINERs identified as unresolved sources at high X-ray energies (4.5--8~keV) suggested that AGN-like objects tend to have higher 2--10 keV luminosity than non-AGN objects, with average values log~L$_{\rm x}$/(\ergs) $\sim$ 41  and log~L$_{\rm x}$/(\ergs) $\sim$ 39, respectively \citep{gonzalezetal09}.
As a further check to firmly identify AGN, we used the X-ray/mid-IR $vs$ X-ray colors relation.\\
In Figure \ref{fig:nhxir_vs_hr4} (left panel) we show the absorption column density $vs$ hardness ratio HR  (HR=(H-S)/(H+S), where H and S are the  source counts in the soft and hard energy bands), for the 24 sources with spectral information. The hardness ratio HR is evaluated using S=2--4.5 keV and H=4.5-10 keV energy bands for \xmm\ and S=0.5-2 keV H=2-8 keV for \chandra.
Sources with larger obscuration are characterized by larger HR. In particular, in our clean sample, a value of HR larger than about $-$0.3 indicates a \nh \gtsima $10^{22}$ cm$^{-2}$ (see Fig. \ref{fig:nhxir_vs_hr4}, left).
\\
Mid-IR emission in AGN is thought to originate within the dusty torus reprocessing the optical-UV primary emission \citep[and references therein]{padovanietal2017}, and mid-IR color--color diagrams offer important diagnostic in order to identify AGN candidates. Using classical color-color diagrams through WISE 3-bands measurements ([2.4-4.6]~$\mu$m  \textit{vs} [4.6-12]~$\mu$m, \citep{mateos2012,assef2013,Yan2013}, our candidates roughly lie the AGN region ([4.6-12]~$\mu$m above 2) and tend toward red colors ([2.4-4.6]~$\mu$m above 0.3), as suggested for AGN in mergers \citep{satyapaletal14,Weston2017}.  Nevertheless, AGN are not the largest population in mid-IR surveys, which are dominated by strong IR emission from SF and normal galaxies. This is the reason why combined mid-IR and X-rays analysis provides further information to confirm AGN candidates \citep{severgnini2012,Terashimaetal2015}. 

\cite{severgnini2012} proposed an effective technique to classify AGN and SF galaxies by comparing mid-IR (12--25 $\mu$m) and X-ray fluxes. In their diagnostic plot, the X-ray/mid-IR flux ratio is compared with the X-ray hardness ratio values, identifying different regions in the plot where different populations (SF galaxies, unobscured AGN, Compton-thin AGN and Compton-thick AGN) are clustered. 
Then we use the ratio F(X-rays)/F(22 $\mu$m) $vs$ HR relation reported in the right panel of Fig. \ref{fig:nhxir_vs_hr4} to exclude X-ray detected targets which are not associated with AGN and evaluate (in Sect. \ref{sect:discussion:absorption}) the level of absorption for the sources in the total sample without X-ray spectral information (see Table \ref{tab:4xmm-sources} and Fig. \ref{fig:nhxir_vs_hr4}, right panel). 
Using the results of spectral analysis for the sources with enough counts (our ``clean" sample), we calibrated the X/mid-IR vs HR plot that allowed  also to identify the regions where different types of emitters are expected (details will be presented in Sect. \ref{sect:discussion:absorption}): we choose the limit HR$>-$0.3 (as obtained for obscured sources with \nh \gtsima $10^{22}$ cm$^{-2}$) and, as X/IR ratio, we adopted a threshold of 0.01 (blue dash-dotted lines in Fig. \ref{fig:nhxir_vs_hr4}).  We note that, given the intrinsic dispersion of the X/mid-IR vs HR relation, these values are not different from those adopted by \cite{severgnini2012}.

Using optical classification (see Fig. \ref{fig:bpt}) and X-ray/mid-IR ratio (see Fig. \ref{fig:nhxir_vs_hr4}, right), we identified and confirmed 42 AGN (4XMM: 33 targets and CSC2: 9 targets), which is 80\% of the detected targets in pairs (52). The last two columns in Table \ref{tab:xfit} report the classification of each source as obtained with Optical and X-ray/mid-IR ratio.
The sources that did not pass at least one of the two checks are indicated in italic in Table \ref{tab:4xmm-sources} and Table \ref{tab:xfit}; however, when available, we report the results of their X-ray spectral fit for completeness. A peculiar but illustrative case is J133817.3+481632 (aka Arp~266) which is a known Compton-thick AGN hosted in a dual system \citep{Iwasawa2020,Mazzarella2012}. It appears as a SF galaxy in our X-ray/mid-IR diagnostic plot and BPT (see src~\# 11 in Fig. \ref{fig:bpt}, lower panel), suggesting that the number of heavily absorbed AGN in our final sample should be consider a lower-limit (see further discussion in Sect. \ref{sect:discussion:fraction}). A similar case of such highly obscured system is NGC~6240 \citep{komossaetal03}.\\
We also note that, for  pair candidates detected by \chandra\ (see Table \ref{tab:4xmm-sources}), four are at angular separation below 12$''$ which corresponds to the WISE 22~$\mu$m resolution \citep{Wright2010}.
One of these four systems is the well known AGN pair Arp~266 that we have just discussed. However, considering the X/mid-IR vs HR diagnostic plot, the decrease of the IR flux (due to blended emission between the two sources) will move the sources in the AGN region, resulting, once again, in a lower-limit in our estimated AGN fraction in the final sample.

%
\begin{landscape}
\begin{table}
\caption{X-ray properties of our clean sample of dual AGN sources detected with \xmm\ and \chandra. Sources in $^\ddagger$\textit{italic} did not pass the check for AGN classification.}
\label{tab:xfit}
\begin{tabular}{|l|l|c|c|c|c|c|c|c|c|c|c|}
\hline
  \multicolumn{1}{|c}{IAUNAME} &
  \multicolumn{1}{c|}{$^{(1)}$\nhgal} &
  \multicolumn{1}{c|}{$^{(2)}$\nh} &
  \multicolumn{1}{c|}{$^{(3)}\Gamma_{\rm h}$} &
 \multicolumn{1}{c|}{$^{(4)}$F$^{\rm Obs}_{\rm X}$} &
  \multicolumn{1}{c|}{$^{(5)}$L$^{\rm Una}_{\rm X}$} &
  \multicolumn{1}{c|}{$^{(6)}$L$_{\rm [OIII]}^{\rm corr}$} &
    \multicolumn{1}{c|}{$^{(7)}$W4} & 
        \multicolumn{1}{c|}{$^{(8)}$E(B-V)} &
                \multicolumn{1}{c|}{$^{(9)}$(X/IR)} &
        \multicolumn{1}{c|}{$^{(10)}$Opt}\\

  \multicolumn{1}{|c}{ (4XMM/2CXO)} &
  \multicolumn{1}{c|}{(10$^{20}$\cm2)} &
  \multicolumn{1}{c|}{(10$^{22}$\cm2)} &
  \multicolumn{1}{c|}{} &
    \multicolumn{1}{c|}{(10$^{-14}$\cgs)} &
  \multicolumn{1}{c|}{(10$^{40}$ \ergs)} &
  \multicolumn{1}{c|}{(10$^{40}$ \ergs)} &
    \multicolumn{1}{c|}{(mag)} &
        \multicolumn{1}{c|}{(mag)} &
                \multicolumn{1}{c|}{}&
                \multicolumn{1}{c|}{}\\
\hline
    \multicolumn{10}{|c|}{XMM-Newton} \\
\hline
   J082321.6+042221 & 2.5 & 0.6$^{+1.4}_{-0.3}$ & 3.03$^{+0.50}_{-0.44}$ & 10.$^{+1}_{-1}$ & 22$^{+4}_{-8}$& >0.015 & 8.4  & - & U & A \\
  
  J100135.8+033648 & 1.9 & 16$^{+31}_{-10}$ & 1.9$^\star$ & 6.3$^{+1.1}_{-1.1}$ & 46$^{+15}_{-15}$ & 21.5 & 6.45$\pm$0.05 & 0.70 & t & A\\
   
   J100236.6+324224  & 1.3 & $<$ 0.4  & 2.2$^{+0.9}_{-0.6}$ & 10$^{+1}_{-1}$ & 47$^{+10}_{-10}$ & - & 8.35 & - & U & C \\
   
   $^\ddagger$\textit{J102142.6+130654 } & 4.3 & $<$ 0.06 & 1.9$^\star$ & 0.8$\pm$0.4 & 10$\pm$4 & >0.06 & 4.34$\pm$0.04 & - & SF & SF\\
   
   $^\ddagger$\textit{J112545.1+144035}  & 2.5 & 0.2$^{+0.4}_{-0.1}$ & 1.9$^\star$  & 1.2$^{+0.2}_{-0.2}$ & 3.3$\pm$1.0 & 13.6 & 3.66$\pm$0.02 & 1.3 & SF & C\\
  J112549.5+144006  & 2.4 & 12$^{+55}_{-8}$ & 1.9$^\star$ & 1.6$\pm$0.3 & 8.0$\pm$0.4 & 6.3 & 4.44$\pm$0.03 & 1.3 & T & C\\
   
   J120443.3+311037  & 1.4 & 3.7$\pm$0.5 & 1.7$\pm$0.4 & 263$\pm$6 & 309$\pm$5 & 43 & 3.59$\pm$0.02 & 1.43 & t & A\\
   J120445.3+311130  & 0.44 & 24$^{+150}_{-12}$ & 1.9$^\star$ & 3.0$^{+1.1}_{-1.1}$ & 11$^{+7}_{-6}$ & >0.2 & 7.92$\pm$0.15 & - & T & A\\
   
   J145627.4+211956  & 2.9 &  75$^{+28}_{-23}$ & 1.9$\pm$0.9 & 8.6$^{+2.0}_{-2.0}$ & 370$^{+20}_{-130}$ & > 4.5 & 7.15$\pm$0.08 & 0.35 & t & A\\
   J145631.3+212030  & 2.9 & $>$ 100& 1.9$^\star$ & 2.3$^{+0.5}_{-0.5}$ & 7
   00$^\dagger$ & 39 & 5.32$\pm$0.03 & 0.54 & T & A\\

    J145838.5+382727  & 1.2 & $<$ 0.6 & 1.9$^\star$ & 1.2$^{+0.2}_{-0.2}$ & 60$\pm$20 & 46.8 & 7.34$\pm$0.08 & 0.63 & t & A\\

    J162640.9+142243  & 3.8 & 67$^{+150}_{-50}$ & 1.9$^\star$ & 2.3$^{+1}_{-1}$ & 80$^{+30}_{-70}$ & 31 & 5.54$\pm$0.11 & 0.44 & T & A \\
    J162644.4+142253  & 3.8 & 6$^{+12}_{-4}$ & 1.9$^\star$ & 1.6$\pm$0.8 & 12$\pm$0.8 & > 0.1 & 9.04$\pm$0.53 & - & t & A\\
    
    J094554.4+423840 & 1.1 & $<$ 0.02 & 2.4$\pm$0.1 & 50$\pm$3 & 700$\pm$30 & 27.9 & 5.45$\pm$0.04 & 0.40 & U & A\\
    J094554.4+423818 & 1.1 & 24$^{+8}_{-6}$ & 1.9$^\star$ & 19$\pm$2 & 690$\pm$ 30 & 124 & 5.84$\pm$0.05 & 0.55 & t & A \\
    
    J103853.2+392151 & 1.7 & $<$ 0.03 & 1.62$\pm$0.07 & 55$\pm$5 & 420$\pm$ 20 & 8.0 & 7.66$\pm$0.14 & 0.33 & U & A \\
    J103855.9+392157 & 1.7 & $>$ 100 & 1.9$^\star$ & 1.6$\pm$1 & 700$^\dagger$ & 19 & 6.05$\pm$0.05 & 0.30 & SF & A \\
\hline
\multicolumn{10}{|c|}{Chandra} \\
\hline
     J090714.4+520343 & 1.5 & 18$^{+16}_{-12}$ & 1.9$^\star$ & 6.4$\pm$1.1 & 160$\pm$30 & 104 & 6.8$\pm$0.1 & 1.11 & t & A\\
     J090714.6+520350 & 1.5 & 2.5$^{+1.4}_{-1.1}$& 1.5$\pm$0.7 & 13$\pm$1 & 160$\pm$1 & 38 & 6.8$\pm$0.1 & 0.85 & t & A \\
     
     J121345.9+024838 & 1.7 & <0.56 & 1.9$^\star$ & 0.7$^{+0.2}_{-0.2}$& 11$\pm$4 & 46 & 3.61$\pm$0.02 & 1.48 & T/SF & L\\
     J121346.0+024841 & 1.7 & -  & 2.1$\pm$0.6 & 0.9$\pm$0.2 & 13$\pm$3 & 80 & 3.61$\pm$0.02 & 1.17 & SF & C\\
     
     $^\ddagger$\textit{J124610.0+304355} & 1.16 & 0.16$^{+0.1}_{-0.8}$& 1.9$^\star$ & 1.1$\pm$0.5 & 1.4$\pm$0.1 & 5.3 & 4.08$\pm$0.03 & 1.37 & SF & SF\\
     J124611.2+304321 & 1.16 & 0.17$^{+0.1}_{-0.1}$ & 1.9$^\star$ & 1.2$\pm$0.3 & 1.5$\pm$0.1 & 0.5 & 7.4$\pm$0.1 & 0.43 & U & C/L\\
     
     \textit{J125934.1+275648} & 0.9 & - & 1.7$^{+0.9}_{-0.1}$ & <0.67 & <1 & 8.4 & - & -\\
     
     \textit{J131513.8+442426} & 1.7 & - & 3.0$^{+1.3}_{-1.1}$ & 0.055 & <0.18 & 150 & 6.19$\pm$ 0.05 & 1.57 & SF& A \\
     J131517.3+442425 & 1.7 & 9.5$^{+0.5}_{-0.4}$ & 1.9$^\star$ & 330$\pm$5 & (1.9$^{-2.1}_{-0.8})\times 10^3$ & 59 & 3.43$\pm$ 0.02 & 0.83 & t & C\\
     
     J133817.3+481632 & 1.4 & >200 & 1.6$\pm$0.3 &  2.2$\pm$0.9 & (1.3$^{-2.7}_{-0.8})\times 10^3$ & 13.8 & 2.54$\pm$ 0.01 & 0.71 & SF & C/SF\\
     J133817.8+481640 & 1.4 & 6.8$\pm$3.8 & 1.9$^\star$ & 24.6$\pm$2.0 & 40. & 14.9 & 2.54$\pm$ 0.01 & - & t & C/SF\\
     
     $^\ddagger$\textit{J150457.1+260058} & 3.5 & - & 1.9$^\star$ &  0.2$\pm$0.1 & 1.5$\pm$0.2 & 0.06 & 8.8$\pm$ 0.1  & - & SF & Amb\\
     $^\ddagger$\textit{J150501.2+260101} & 3.5 & - & 1.1$^{+0.9}_{-0.7}$ & <0.1 & <0.9 & 0.08 & 8.8$\pm$ 0.1 & - & SF & Amb \\
     
     $^\ddagger$\textit{J154403.4+044607} & 3.9 & - & 2.9$^{+1.2}_{-1.1}$ & <0.1 & <0.46 & 150 & 6.38$\pm$0.07 & 2 & SF & L\\
    J154403.6+044609 & 3.9 & 2.8$^{+2.8}_{-1.9}$ & 1.3$^{+1.2}_{-1.0}$ & 8.1$\pm$0.9 & 48$\pm$9 & 25 & 6.38$\pm$0.07 & 1.2 & t & L\\
    \hline
\end{tabular}\\
$^{(1)}$ Galactic column density of NI from \citealt{HI4PI}; $^{(2)}$ Nuclear absorption column density. "-" refers to source with upper limit on \nh\ of fews 10$^{20}$ \cm2; $^{(3)}$ Photon index of the hard power-law; values with $^\star$ are left fixed in the fit; $^{(4)}$ Observed 0.5--10 keV (4XMM) and 2--8 keV (2CXO) X-ray flux in 10$^{-14}$\cgs; $^{(5)}$ Unobscured 2-10 keV X-ray Luminosity in 10$^{40}$\ergs; $^\dagger$ For Compton-thick sources, we used the relation in \cite{lamastraetal09,marinuccietal12}: the intrinsic 2–10 keV luminosity has been obtained by multiplying the observed luminosity by 70; $^{(6)}$ De-reddened [OIII] luminosity in 10$^{40}$\ergs; $^{(7)}$ WISE W4 magnitude.; $^{(8)}$ Extinction considering the Balmer decrement from narrow emission lines. "-" indicates sources where precise measure of H$\beta$ has not be obtained (see Sect. \ref{sect:opt analysis} for details; $^{(9-10)}$ Source classification following the X/IR vs HR  and Optical diagnostics. A: AGN, SF: Star-Forming, L: Liner, C: Composite, U: Unabsorbed, t: Compton thin, T: Compton Thick, Amb: Ambiguous.
\end{table}
\end{landscape}

\section{Discussion}
\label{sect:discussion}

\subsection{Absorption properties}
\label{sect:discussion:absorption}

We were able to extract X-ray spectra and perform a detailed analysis for 32 sources among the 52 AGN candidates in 4XMM and CSC2.
The optical and X-ray/mid-IR presented in Sect. \ref{sect:AGN ident} points towards an AGN classification for 24 out of 32 targets.
We present the spectral analysis for all 32 sources with best-fit values in Table \ref{tab:xfit}; however, we excluded from our final considerations on dual-AGN properties all sources (8) which are not confirmed as AGN (reported in italic in Table \ref{tab:xfit}). 

\begin{figure*}
\centering
 \includegraphics[width=0.45\textwidth]{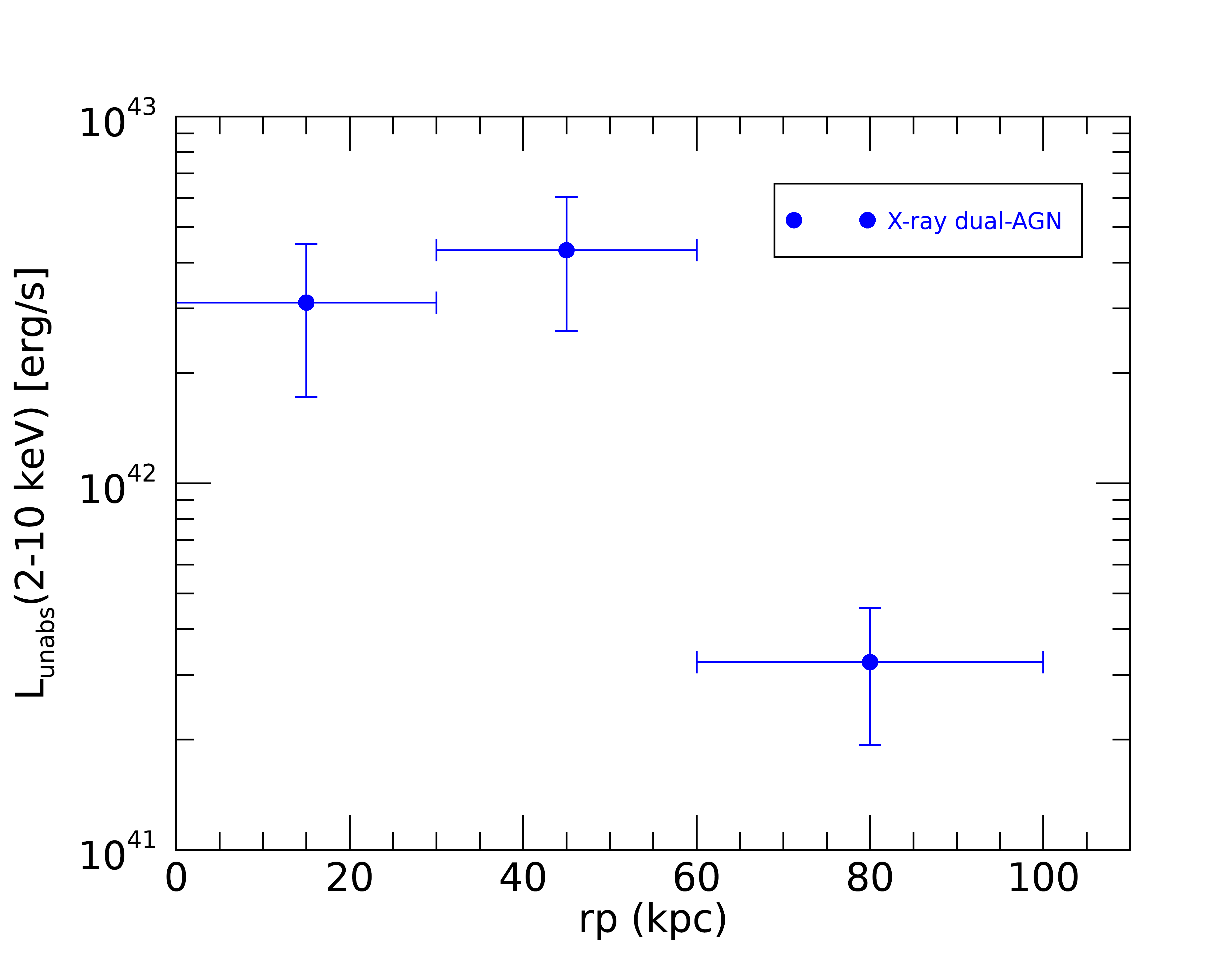} 
\includegraphics[width=0.45\textwidth]{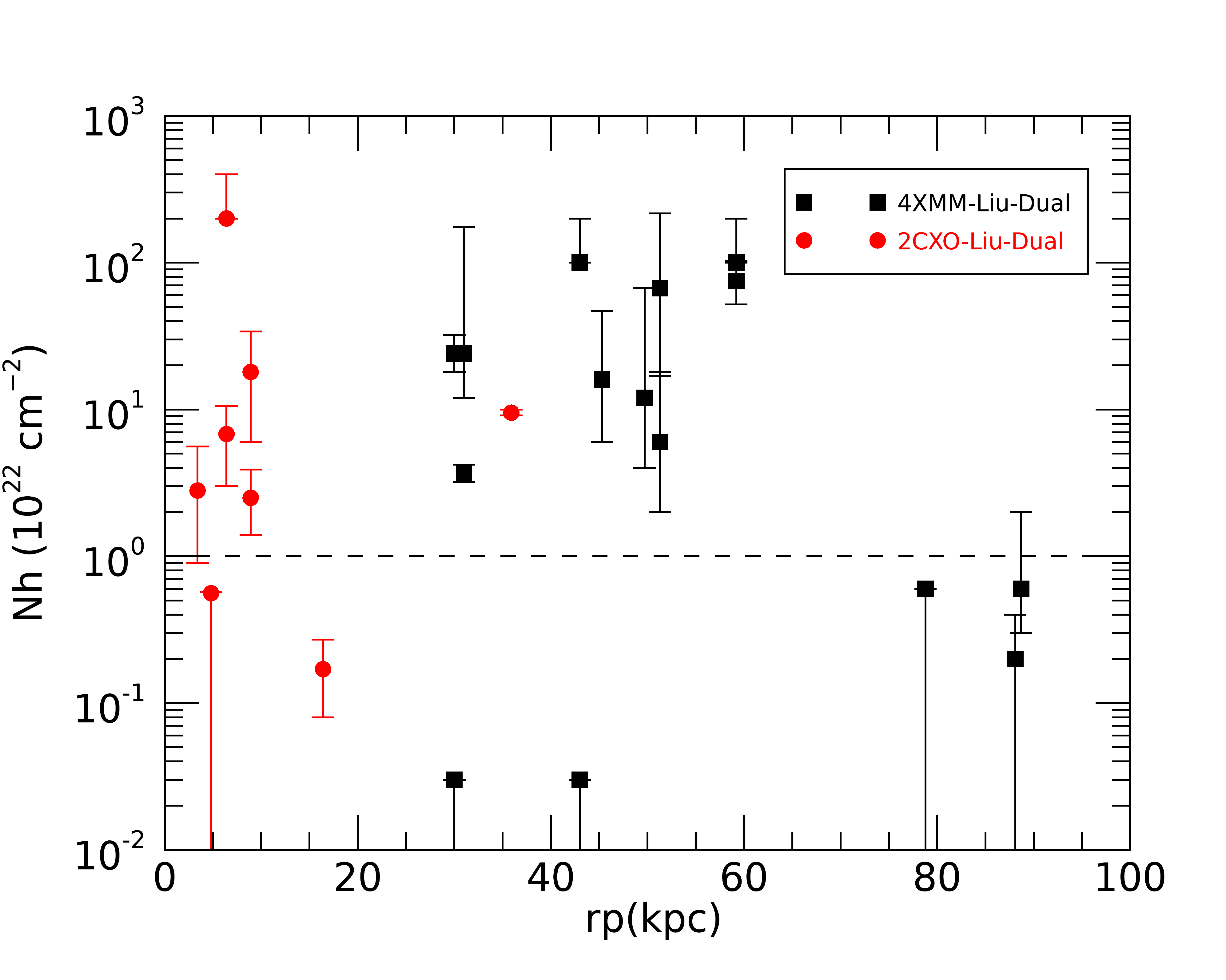}
\caption{Broad-band intrinsic X-ray luminosity (left) and absorbing column density N$_{\rm H}$ (right) $vs$ projected separation rp for clean sample of X-ray dual AGN detected in 4XMM and CSC2 and with spectral information (see Table \ref{tab:xfit}). L$_{\rm unabs}$ errors are dominated by dispersion in each single bin.} \label{fig:Lnhvs_rp}
\end{figure*}

In the left panel of Figure~\ref{fig:Lnhvs_rp}, we plot the intrinsic unobscured X-ray luminosity of our dual-AGN sample  with spectral information in bins of projected separation rp. A trend of increasing luminosity with decreasing rp is suggested, although it should be considered that the bin rp$>$ 60~kpc is poorly populated (3 AGN) in this analysis, while smaller separation bins are populated with 10 (0--30~kpc) and 11 (30--60~kpc) targets.

Among the 24 targets with measured \nh, a fraction of 70--74\% exhibits an absorption column density larger than 10$^{22}$ cm$^{-2}$ at 3--100~kpc separation (see the right panel in Figure \ref{fig:Lnhvs_rp}). 
We measure an upper limit  N$_{\rm H}<$3$\times$20 cm$^{-2}$ for the two AGN showing  broad optical emission lines (J094554.4+423840 and J103853.2+392151, \citealt{derosaetal18} see Sect. \ref{sect:opt analysis}).
Although we do not see a clear trend of absorption with rp (but see further discussion below based on the absorption on the whole sample), we note that there are not absorbed AGN (N$_{\rm H}>$10$^{22}$ cm$^{-2}$) in systems with projected separation above 60~kpc. The fraction of absorbed AGN in dual systems is larger than that measured in samples of isolated AGN (e.g., 45\% in BAT, \citealt{riccietal15}), which is in agreement with previous studies on X-ray dual AGN  \citep{kossetal11,derosaetal18,riccietal17,riccietal2021}.  However, we note here that our sample is neither complete nor unbiased in any sense. In order to evaluate the amount of absorption in the 42 X-ray dual AGN, we use indirect measurements. 
Below we describe two main proxies for absorption: [OIII]/X ratio, IR/X ratio $vs$ hardness ratio.\\

\subsubsection{L$_{\rm x}$ $vs$ L$_{\rm [OIII]}$}

Once that extinction within the NLR is properly considered, the luminosity of the emission line from ${[\rm OIII]\lambda}$5007  may be used as a good indicator of the intrinsic luminosity of the source.
In our analysis of optical spectra (Sect. \ref{sect:opt analysis}), we correct the [O\ III] emission for the extinction through the Balmer decrement (H$\alpha$/H$\beta$ ratio).
To derive the L$_{\rm [OIII]}^{\rm corr}$ corrected for extinction, we used the relation from \cite{bassanietal99} which assumes the \cite{cardellietal89} extinction law and an intrinsic Balmer decrement equal to 3; this value represents the case for the NLR \citep{osterbrock&ferland06}. Value of L$_{\rm [OIII]}^{\rm corr}$ for the sources with X-ray spectral information are reported in Table \ref{tab:xfit}.
The ratio between the observed X-ray luminosity and the L$_{\rm [OIII]}^{\rm corr}$ has been therefore used as an indirect measurement of \nh.
The L$_{\rm x}$/L$_{\rm [OIII]}^{\rm corr}$ relation has been deeply analysed in large samples of type~1 and type~2 AGN \citep{heckmanetal05, mulchaeyetal94, bassanietal99, lamastraetal09, vignalietal10}, and empirical trends have been found by several authors.

\begin{figure}
\includegraphics[width=0.45\textwidth]{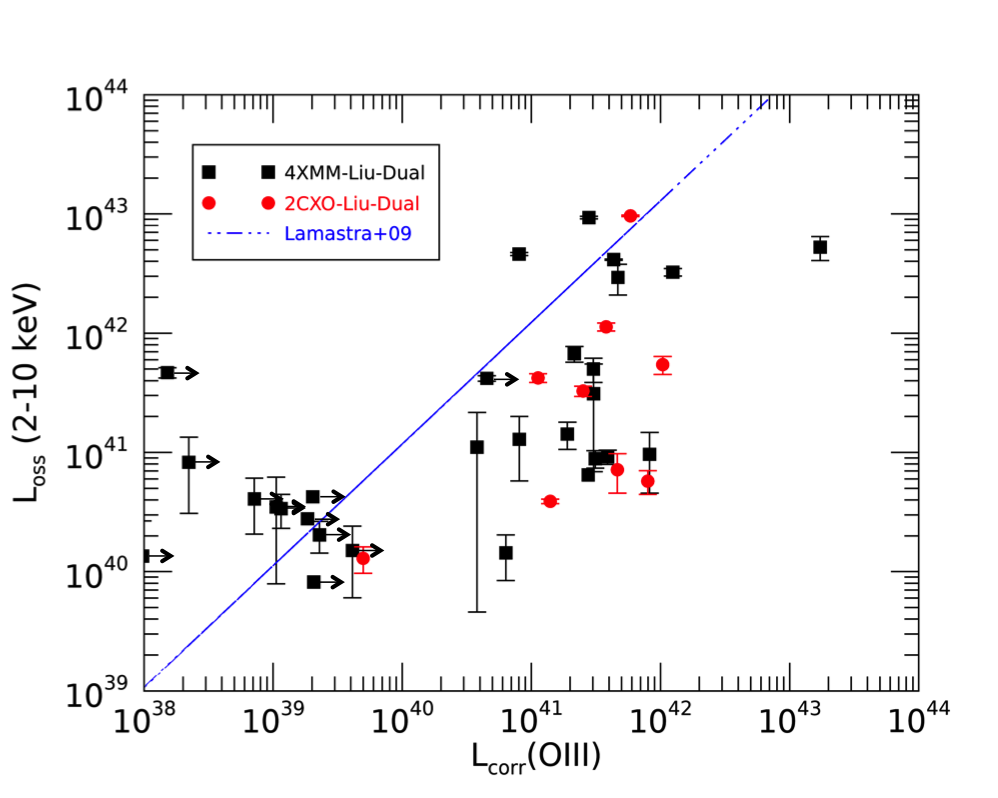} 
\caption{X-ray observed luminosity $vs$ de-reddened [OIII] luminosity for the whole sample of confirmed AGN. 4XMM: black squares, CSC2: red dots. Lower limits to L$_{\rm [OIII]}^{\rm corr}$ refer to the sources without a measure of H$\beta$ emission, for those we use the observed value of [O\ III]. The blue dot-dashed line represents the relation found in \citealt{lamastraetal09} for a sample of  Compton-thick AGN. All sources in pairs lie below the expected relation, suggesting a larger absorption with respect to isolated AGN. The two type~1 AGN in the 4XMM sample are the only measured data points above the relation.} \label{fig:Lx_Loiii}
\end{figure}

In fig. \ref{fig:Lx_Loiii}  we report the observed X-ray luminosity $vs$ the [O\ III] luminosity corrected for extinction L$_{\rm [OIII]}^{\rm corr}$.
For the sources without a measure of H$\beta$ emission, we considered the observed value of [O\ III] that should be then regarded  as a lower limit to L$_{\rm [OIII]}^{\rm corr}$. As comparison, we also plot the relation found by \cite{lamastraetal09} for a sample of Compton-thin AGN. 
Almost 80\% of our sample lies below the threshold as defined in \cite{lamastraetal09}, clearly showing that our sources exhibit excess of absorption with respect to what is expected in isolated Compton-thin AGN. In the X-[O\ III] plane, the  two type~1 AGN in our sample (094554.4+423840 and J103853.2+392151) are both located above the relation defined for obscured sources, as expected. 

With the caveat that for some sources of the sample only a lower limit to the H$\beta$ emission has been obtained, we can compare the column density derived from the E(B-V) -- evaluated from the NLR (see Table \ref{tab:xfit}, \citealt{Domingues2013}) and using, for the conversion, the Galactic N$_{\rm H}$/E(B-V) ratio \citep{Draine2011} -- with the N$_{\rm H}$ measured from the X-ray analysis (excluding type~1 AGN). The extinction in the NLR is lower than the obscuration derived from X-rays, and there is no evidence for a trend of increasing E(B-V) as a function of the separation rp as observed for X-ray derived N$_{\rm H}$ in Fig. \ref{fig:Lnhvs_rp} (right panel). This result suggests that the gas responsible for the X-ray obscuration is likely associated with the torus and/or the BLR (e.g.,  \citealt{padovanietal2017}).

\subsubsection{F$_{\rm x}$/F$_{\rm [IR]}$ $vs$ X-ray hardness ratio}

As anticipated in Sect. \ref{sect:AGN ident}, we used mid-IR/X-ray flux ratio to identify the AGN among the X-ray detected targets (see Table \ref{tab:4xmm-sources}). This ratio, when compared with X-ray colors, provides a powerful diagnostic for obscuration \citep{severgnini2012}, also when \nh\ is not available from X-ray fitting procedures.
We then use the clean sample with spectral information (24 AGN, see Table \ref{tab:xfit}) to calibrate the \nh\ vs HR relation to be used in the total sample, as discussed in Sect. \ref{sect:AGN ident} and shown in Fig. \ref{fig:nhxir_vs_hr4} left panel. 

The plot X/mid-IR vs HR in Fig. \ref{fig:nhxir_vs_hr4} right allow us to identify four regions for unabsorbed, SF galaxies, Compton-thin and Compton-thick (CT) sources. 
The sources in the bottom-left block are identified as star-forming galaxies also from optical analysis (see Sect. \ref{sect:opt analysis} and Fig.~\ref{fig:bpt}).
We checked that, within the diagnostic plot, all sources analysed through X-rays and optical analysis  are correctly located: in Table \ref{tab:xfit} we report the identification obtained with all diagnostic, X-rays, optical and X/MIR $vs$ HR. We, however, noted that  J133817.3+481632 (a.k.a. Arp~266 SW) and J103855.9+392157, which are Compton thick, fall in the SF region, being their spectra dominated by the soft X-ray component with the nuclear emission completely obscured \citep{Iwasawa2020,derosaetal18}. This is also in agreement, as anticipated in Sect. \ref{sect:AGN ident}, with the optical classification of these targets, yet demonstrating that we can miss a number of Compton-thick sources in our final sample of confirmed AGN.

If we consider all 42 confirmed AGN detected in 4XMM (33) and CSC (9) in fig. \ref{fig:nhxir_vs_hr4} (right panel), the mid-IR and X-ray diagnostic suggests that our sample is composed of 80\% Compton-thin AGN, 16\% CT and 4\% unobscured sources. As mentioned above, the number of CT sources can be regarded as a lower-limit. If we compare these fractions to isolated AGN, we confirm the trend of more obscured AGN hosted in pairs. As pointed out above, we do not see a clear trend of increasing \nh\ with decreasing rp (see Fig. \ref{fig:Lnhvs_rp}, left). However, the fraction of absorbed AGN in these dual-AGN systems that are in early stages of merger (projected separation up to 100 kpc) is lower with respect to the fraction of heavily obscured AGN found in systems in late stages of merging (e.g. \citealt{riccietal2021}), suggesting the presence of larger reservoir of nuclear gas or with a different distribution in AGN at closer separation. 

We note that, as described in the previous section, the bulk of absorption we measure is likely associated with nuclear scale (BLR/torus). This indicates that the observed systems are in environments where gas is transported closer to the AGN and/or disturbed in dual AGN separated by <60\,kpc. Based on the data discussed in this work, we find no strong evidence that AGN pairs with projected separations in the $\sim 1-100\,$kpc range prefer gas-rich galaxies (i.e., with a larger fraction of gas mass than other galaxies;  \citealt{Kewleyetal06}). If confirmed in subsequent observations, this would imply that interactions with gas (via tides, spiral arms, dynamical friction, etc.) are not a necessary mechanism for orbital evolution of AGN pairs from 100\,kpc scales. Samples like the one analyzed in this paper could therefore be used to place some initial observational constraints on theoretical models describing orbital evolution of MBH pairs.

\subsection{X-ray detection efficiency of dual-AGN candidates}
\label{sect:discussion:fraction}

In this section we report the study to evaluate the efficiency of X-ray observations in order to detect optically selected dual-AGN candidates. To this aim, we considered the optically selected sample of AGN pairs and  measure the ratio between the number of detected optical AGN and the number of observed optical AGN along the \xmm\ and \chandra\ pointings, Xeff=DET/OBS.
To increase the available statistics, in addition to the sample of AGN pairs from \citealt{liuetal11}, we  considered further samples of dual AGN identified from both optical image \citep{Mezcua2014} and double-peaked emission-line techniques \citep{Ge2012,Wang2009,Smith2010,Kim2020}.
The list of the selected samples are reported in table \ref{tab:fractions2}.
We note that almost 90\% of X-ray detected targets in all samples have been observed more than 10~ks of elapsed time.\\
Moreover, most of the detected sources have been observed serendipitously, with an off-axis position up to 12$'$--15$'$.
In order to select the number of observed AGN candidates, we consider here two thresholds: exposure and off-axis position.
To take into account the fastest drops of \chandra\ sensitivity at increasing  off-axis angles with respect to \xmm, we selected all targets in the \xmm\ and \chandra\ field of view with an off-axis position lower than 15$’$  and 10$’$, respectively and with exposure larger than 10~ks. Of course, the number of observed AGN might depend of the threshold we assumed; however, we verified that the number reported below are stable against other choices of off-axis position and exposures close to the selected ones.\\
In Table \ref{tab:fractions2} we report the ratio Xeff of all the optically selected dual AGN detected in 4XMM and CSC2, a row for each optical sample we checked.
We stress that these detection fractions should be consider as lower-limits for two main reasons. 
For what concerns the numerator of the ratio Xeff (the X-ray detected optically selected AGN), we note that the upper limit for the undetected sources is in the range  0.004--0.04 cts/s (3$\sigma$ c.l.) in the 4XMM 2--12 keV band, suggesting that we may fail to detect heavily obscured sources  with X-ray luminosity lower than 10$^{42}$\ergs, considering an average redshift for our sample of z=0.03 and a standard spectral slope of $\Gamma$=2.
About the denominator, the X-ray observed optically selected AGN, another important point to consider is the possibility that the optical classification as AGN in the samples under investigation could be instead due SF emission in the galaxy. In fact, as we showed through our MWL analysis (see Sect. \ref{sect:AGN ident}), about 80\% of the optically selected targets are confirmed as AGN. If we consider this fraction, the X-ray detection efficiency considering all optically selected dual AGN in 4XMM and CSC2 (last column in Table \ref{tab:fractions2}) is about 60\%.
Increasing the exposure cut to 15 ks, the detection efficiency weakly increases from 62 to 85\%. This suggests that, when observed properly, X-ray data represent a powerful technique to confirm and investigate dual-AGN systems.
In this regard, we note that in the sample with higher X-ray detection fraction (\citealt{Kim2020}, \citealt{Mezcua2014} and \citealt{Ge2012} for 4XMM, \citealt{Wang2009}, \citealt{Smith2010} and \citealt{Kim2020} in the case of \chandra), the average value of the off-axis position of the observed targets is lower than in the sample with low X-ray detection efficiency.

\begin{table*}
\caption{X-ray detection efficiency of optically selected dual AGN. This is the ratio between the detected optically selected AGN and the observed ones in the field of \xmm\ and \chandra\ observations.}
\label{tab:fractions2}
\begin{tabular}{|l|l|l|l|l|}
\hline
\multicolumn{1}{|l|}{Catalogue} &
  \multicolumn{1}{|l}{$^{(1)}$Chandra} &
    \multicolumn{1}{l|}{$^{(1)}$XMM} &
   \multicolumn{1}{l|}{X-ray tot (\%)} \\
   \multicolumn{1}{|l|}{} &
  \multicolumn{1}{|l}{det/obs (\%) } &
    \multicolumn{1}{l|}{det/obs (\%)} &
   \multicolumn{1}{l|}{det/obs (\%)} \\
    \hline
    \multicolumn{4}{|c|}{T>10 ks and R$_{\rm offset}$<15$'$(XMM)--10$'$(Chandra)} \\
\hline
\citealt{liuetal11}     & 51/137 (37) & 73/165 (44) & 124/302 (41) \\
\citealt{Mezcua2014}    & 5/16 (31) & 7/8 (88) & 12/24 (50) \\
\citealt{Ge2012}        & 11/18 (61) & 15/17 (88) & 26/35 (74)\\
\citealt{Wang2009}      & 8/10 (80) & 2/5 (40) & 10/15 (67)\\
 \citealt{Smith2010}    & 14/17 (82) & 6/11 (55) & 20/28 (71)\\
 \citealt{Kim2020}      & 11/11 (100) & 5/6 (83) & 16/17 (94)\\
$^{(2)}$Tot         & 100/209/0.8 (60) & 108/212/0.8 (64) & 208/421/0.8 (62)\\
\hline
\hline
    \multicolumn{4}{|c|}{T>15 ks and R$_{\rm offset}$<15$'$(XMM)--10$'$(Chandra)} \\
\hline    

$^{(2)}$Tot           & 100/154/0.8 (81) & 108/152/0.8 (89) & 208/306/0.8 (85)\\
 \hline \end{tabular}\\
 $^{(1)}$ Fraction of AGN detected in 4XMM/CSC2 catalogues over observed with more than 10 ks (upper raws) and 15~ks (last raw)  considering an off-axis position <10$'$ for \chandra\ and <15$''$ for \xmm. Exposure time refers to elapsed time; $^{(2)}$ The total has been evaluated considering that the number of confirmed AGN  is the 80\% of the observed targets (see details in Sect. \ref{sect:discussion:fraction}).
\end{table*}

\section{Conclusions}
\label{sect:conclusions}
We have investigated the properties in the X-ray domain of a sample of optically selected dual AGN with projected separation between 3--97 ~kpc. Using optical, mid-IR and X-ray diagnostic tools, we were able to characterize the intrinsic properties of this sample and compare them with those of isolated AGN.
\begin{itemize}

\item Among 124 X-ray detected dual-AGN candidates, 52 appear in pairs and 72 as single X-ray AGN. We focused our study on AGN selected in pairs, while AGN detected in single sources will be analysed in a forthcoming paper (Parvatikar et al., in preparation).
\item Using optical spectroscopy (BPT diagrams in Fig. \ref{fig:bpt}) and X/mid-IR $vs$ X-ray HR (Fig. \ref{fig:nhxir_vs_hr4}), we confirmed as X-ray dual AGN 42 (80\%) sources (either LINER or AGN). The X-ray luminosity of LINER sources strongly favours a scenario in which the source emission is accretion-driven. Due to possible identification of heavily obscured AGN with SF (see e.g. case of J133817.3+481632  and J103855.9+392157) and the blended IR emission of unresolved pairs in W4 band, this fraction should be considered as a lower-limit.

\item We confirm the trend of increasing AGN luminosity with decreasing separation (see left panel of  Fig.~\ref{fig:Lnhvs_rp}),  suggesting that mergers may trigger more luminous AGN.
\item When comparing optical (de-reddened [O~III]) and observed X-ray luminosities (i.e., not corrected for the obscuration; see Fig.~\ref{fig:Lx_Loiii}), our sample shows, on average, a larger  obscuration with respect to the relation found for obscured Seyfert galaxies in isolated systems. Moreover, systems at closer separation show a higher obscuration with respect to dual AGN with separations rp above 50--60 kpc.
\item The N$_{\rm H}$ measured from X-ray spectral analysis is always higher than the absorbing column density derived form the extinction E(B-V) evaluated for NLR, suggesting that the gas responsible from the obscuration should lie in nuclear regions (likely the torus or BLR). 
\item Using X/mid-IR ratio $vs$ HR (see Fig. \ref{fig:nhxir_vs_hr4}), we estimate that a fraction of 80\% of the confirmed AGN are Compton thin (with N$_{\rm H}$ larger than 10$^{22}$ \cm2) and 16\% are Compton thick (N$_{\rm H}$ larger than 10$^{24}$ \cm2). These fractions are larger if compared with samples of isolated systems, but lower with respect to the fraction of heavily obscured AGN found in systems in late-stage of the merging process (projected separation below 10--20 kpc). This evidence is suggesting that pairs of AGN are more heavily obscured with respect to isolated AGN.

\item These findings indicate that dual AGN find themselves in environments where gas is transported closer to the AGN and/or disturbed for dual AGN separated by <60\,kpc. If the host galaxies of the dual AGN in this sample are further showing to contain similar amounts of gas as galaxies with single AGN, this would suggest that interactions with gas are not a necessary mechanism for orbital evolution of AGN pairs from 100\,kpc scales.
\item When different samples of dual AGN are considered, we found that the X-ray detection efficiency (defined as the ratio between the X-ray detected and observed optical AGN in dual systems) lies in the range 62--85\%, depending on the exposure we considered (10--15~ks, see Table \ref{tab:fractions2}).
This fraction should be considered as a lower-limit because, due to the limited exposures of X-ray dual systems (mainly observed serendipitously) in 4XMM and CSC2 catalogues, we miss in our analysis all the heavily obscured AGN with 2--10 keV luminosity lower than 10$^{42}$ \ergs\ at z=0.03.

\end{itemize}

\section*{Acknowledgements}
We would like to thank the referee for the careful reading of the manuscript and useful comments that helped to improve its quality.
We acknowledge financial contribution from Bando Ricerca Fondamentale INAF 2022 Large Grant  “Dual and binary supermassive black holes in the multi-messenger era: from galaxy mergers to gravitational waves”.
ADR, PS, CV, SB acknowledge financial contribution from the agreement ASI-INAF n.2017-14-H.O.
MC acknowledges support from NSF AST-200793.
This research has made use of data obtained from the 4XMM \xmm\ serendipitous source catalogue compiled by the 10 institutes of the \xmm\ Survey Science Centre selected by ESA. This research has made use of data obtained from the \chandra\ Source Catalog, provided by the \chandra\  X-ray Center (CXC) as part of the \chandra\ Data Archive. 
This publication makes use of data products from the Wide-field Infrared Survey Explorer, which is a joint project of the University of California, Los Angeles, and the Jet Propulsion Laboratory/California Institute of Technology, and NEOWISE, which is a project of the Jet Propulsion Laboratory/California Institute of Technology. WISE and NEOWISE are funded by the National Aeronautics and Space Administration.
Funding for SDSS-III has been provided by the Alfred P. Sloan Foundation, the Participating Institutions, the National Science Foundation, and the U.S. Department of Energy Office of Science. The SDSS-III web site is http://www.sdss3.org/.
SDSS-III is managed by the Astrophysical Research Consortium for the Participating Institutions of the SDSS-III Collaboration including the University of Arizona, the Brazilian Participation Group, Brookhaven National Laboratory, Carnegie Mellon University, University of Florida, the French Participation Group, the German Participation Group, Harvard University, the Instituto de Astrofisica de Canarias, the Michigan State/Notre Dame/JINA Participation Group, Johns Hopkins University, Lawrence Berkeley National Laboratory, Max Planck Institute for Astrophysics, Max Planck Institute for Extraterrestrial Physics, New Mexico State University, New York University, Ohio State University, Pennsylvania State University, University of Portsmouth, Princeton University, the Spanish Participation Group, University of Tokyo, University of Utah, Vanderbilt University, University of Virginia, University of Washington, and Yale University. 

\vspace{-0.2cm}
\section*{Data Availability}
The high-level data underlying this article are extracted through
standard processing from raw data stored in public archives (SDSS, \xmm, \chandra\ and WISE), and will be shared on reasonable request to the corresponding author.

\bibliographystyle{mnras}
\bibliography{magna_library}

\label{lastpage}
\bsp	
\end{document}